\documentclass[aps,prd,showpacs,eqsecnum,amsmath,amssymb,nofootinbib,onecolumn]{revtex4}
\usepackage[utf8]{inputenc}
\usepackage{graphicx}
\usepackage{subfigure}
\usepackage{array}
\usepackage{multirow}
\usepackage{amsmath}
\usepackage{amssymb}

\newcommand{\compcent}[1]{\vcenter{\hbox{$#1\circ$}}}

\newcommand{\comp}{\mathbin{\mathchoice
  {\compcent\scriptstyle}{\compcent\scriptstyle}
  {\compcent\scriptscriptstyle}{\compcent\scriptscriptstyle}}}

\usepackage{hyperref}
\usepackage{cleveref}
  
\begin{document}

\title{Gauge independent response of a laser interferometer to gravitational waves.}
\author{Arkadiusz B{\l}aut}
\affiliation{Institute of Theoretical Physics, University of Wroc\l aw, Wroc\l aw, Poland}

\begin{abstract}
Laser interferometer response to a plane gravitational wave on the Minkowski background is given. The derivation does not assume any particular gauge within a class compatible with almost Minkowskian coordinates that preserve a plane wave form of the solutions. Consequently all ten modes of the metric are taken into an account. The final result, the time of flight of a laser signal exchanged between freely moving observers, is expressed in the form of integrals of the metric perturbations taken along the undisturbed trajectories. The result can be applied in any metric theory o gravity; it is valid in the first order of the perturbation and does not assume the long wavelength approximation. The obtained response is shown to define an observable, i.e. it is gauge--invariant with respect to the assumed class of gauge transformations. 
\end{abstract}


\maketitle

\section{Introduction}

To detect gravitational waves the detailed analysis of the detector response is particularly
important. It is usually understood that an incoming gravitational wave affects the motion of freely falling test masses giving an opportunity to measure their time--varying relative distances. 
On the other hand an alternative statement is also put forward that a measurable effect arises because a passing gravitational wave modifies the paths of the laser beam stretching and squeezing its wavelength along the way between an emitter and a receiver. These two descriptions without further qualifications can lead to confusions and misinterpretations. First, they can be read as coordinate dependent statements, in which case they partially characterize the interaction between the gravitational wave and measuring apparatus, and partially just carry an information on the nature of coordinates used \cite{MTW}. Second, expressions as 'affects the motion' or 'modifies the path' demand answering the preceding question, to which standard of rest or uniform motion they refer. In turn the statement 'stretches and squeezes the wavelength' assumes some state of motion of a local oscillator that is to be used in the relevant measurement. The origin of these difficulties can be traced back to a fundamental feature of the Einstein theory, that is, its geometric and relational character. It demands that any measurable quantity should be based on relations between dynamical, geometrical objects. The point is that the descriptions that relay on coordinates are of course allowed and in most cases unavoidable but they are to some degree conventional, and moreover, one must assure that the final result is coordinate independent. 

The most common approach in analyzing the response of the detectors or, more generally, to study various effects of gravitational field, starts with a choice of a preferred coordinate system. For laser interferometrs operating on the Earth (see e.g. \cite{AdLIGO}) particularly useful is the local Lorentz frame (LLF) of the emitter or, in this case, the beam splitter. It manifests its advantage especially when the long wavelength (LW) approximation is valid, i.e. when the wave of the incoming signal is longer than the detector's arm (\cite{LG77}, \cite{GP80}). Then a coordinate dependent description is justified according to which passing gravitational wave put into motion the end mirrors while undisturbed laser light serves merely as a ruler.
Here 'undisturbed' means that in these coordinates the world lines of light rays are stright lines inclined on 45${}^\circ$. An extension of the LLF beyond the leading LW order was elaborated in \cite{BG04}; the LW restrictions were next fully overcame in the case of a plane gravitational wave in \cite{MR14}, where the use of other closely related coordinates (optical coordinates and wave-synchronous coordinates) was advocated as well. This is important since in the future Earth--based inteferometers (like Einstein Telescope \cite{ET}) going beyond the LW approximations seems to be inevitable.
Usually, to avoid the LW restriction one employs the transverse-traceless (TT), or synchronous, coordinates \cite{MTW} which are especially convenient in studies of interferometric missions in space, such as the most advanced  project LISA \cite{LISA}. This future gravitational waves laboratories are dedicated to observations of multitude of sources in a wide frequency band (from a fraction of millihertz to a fraction of hertz), including those with wavelengths 
comparable to the distance between spacecraft (of the order of a ten seconds). In TT coordinates the world lines of $x, y, z =$ const. are timelike geodesics, therefore the coordinate positions of a freely falling fleet of spacecraft (approximately) maintain their unperturbed tracks, and only the optical paths of light become modified. 

Furthermore, a detailed knowledge of the detector response is crucial in testing alternative theories of gravity (for a review see \cite{WillBook}--\cite{YS13} and references therein). The competing theories predict  gravitational waves in all possible polarization states (see \cite{Eardley73} for the pioneering work regarding the metric theories of gravity at the time, and e.g. \cite{PWBook}--\cite{IWMP15} for a model independent approach to study detection capabilities of the present or near future missions) moving with possibly different velocities, as is the case for the Einstein--Aether theory \cite{Jac07}, \cite{YBBY17}, higher order theories \cite{BCL10}, or the so called $f(R)$ theories \cite{CST10}. With this motivations the interferometric detector response for the scalar waves, for massless and massive modes, was studied in some details in \cite{MagNic00}--\cite{AB2015} in the conformal coordinates. In this gauge, which is not restricted by the LW limits, light rays remain undisturbed while timelike geodesics wiggle under the influence of a gravitational wave.

To elucidate the geometric content of the detection process which may be hidden when working in a specific gauge, a coordinate--free description would be highly welcome. Fully geometric account of the detector response was achieved recently in \cite{KF14} for a general gravitational field and in the context of gravitational waves. The observed evolution of the clock phase recorded by the detector with the use of the laser signals was expressed in terms of the Riemann curvature tensor integrated along the light trajectories.

In the present paper we analyze the laser interferometer response to an arbitrary plane gravitational wave on the Minkowski background. In the proposed approach we explicitly make use of coordinates, and from the beginning we are dealing with the gauge dependent concept of the metric perturbation. As a basic measurement we consider the flight time of the light between two freely moving observers. (From this one can derive other observables, like the Doppler tracking signal.) The final result, the time of flight, is expressed in the form of integrals of the metric perturbations taken along the undisturbed trajectories. The derivation, although uses coordinates, differs from most of the previous coordinate dependent approaches in that it does not assume any particular gauge. Consequently, we are forced to consider all ten perturbation modes. Some of them may be spurious signals arising from the use of specific coordinates, i.e. they may represent "coordinate waves". Of course, the final response must be sensitive only to true degrees of freedom. To show that this is the case the gauge invariance of the final result with respect to a class of suitable gauge transformations is demonstrated. These gauge transformations act within the space of plane wave solutions with a fixed propagation vector; this condition specifies the form of gauge generators, cf. Sec.\ref{sec:gi}.

The paper is organized as follows. In the Section \ref{sec:eom} we solve the equations of motion for the observers and light signal. In Section \ref{sec:dtau} we define the basic measurement and we derive our main result, the formula for the time of flight. Section \ref{sec:num} compares the analytic solution obtained in the preceding section with the numerical example for a specific experimental setting. In Section \ref{sec:gi} we explicitly show that the final result is gauge invariant.

Notation and conventions: here and in what follows we use almost Minkowski coordinates $x^{\alpha}$, $\alpha=0,1,2,3$ (we assume $c=1$): $x^{}=(t,{\mathbf x})$, 
${\mathbf x}=(x^1,x^2,x^3)$, $x^0=-x_0=t, x^1\equiv x, x^2\equiv y, x^3\equiv z$, $A^0\equiv A^t, A^1\equiv A^x, A^2\equiv A^y, A^3\equiv A^z$, etc.; in these coordinates metric has the form ${\rm g}_{\alpha\beta}=\eta_{\alpha\beta}+{\rm h}_{\alpha\beta}$, $\;\eta=\text{diag}(-,+,+,+)$, $\;{\rm h}_{\alpha\beta}\ll 1$.

We also add a comment regarding the terminology. We adopt the language in which gravitational waves are treated as small perturbations of the Minkowski background; slightly abusing the terminology we will call them perturbations even if they appear as a result of curvilinear coordinates on the flat spacetime.

\section{Equations of motion for the observers (emitter and detector) and the light ray}
\label{sec:eom}

\begin{figure}[htp]
\begin{center}
\includegraphics[width=19pc]{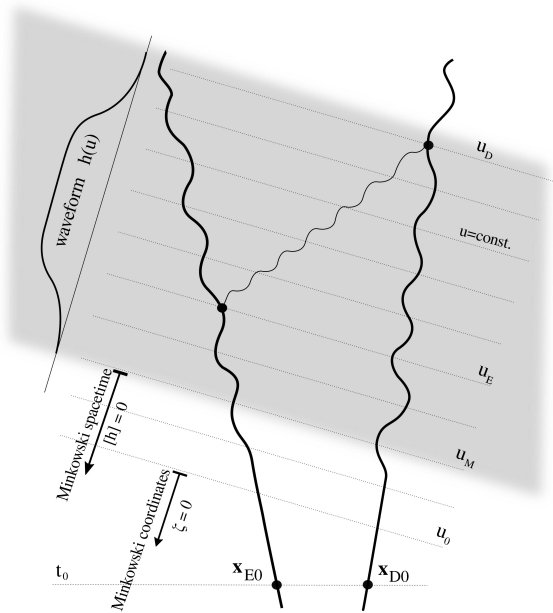}
\end{center}
\caption{Spacetime diagram of the experiment. The world lines of observers (thick lines) and light ray (thin line) are depicted. Initial conditions are fixed (with respect to Minkowski coordinates) before the wave arrival.}
\label{fig:1}
\end{figure}

In general relativity as well as in other geometric theories of gravity the notion of the distance must be precisely defined. In the laser interferometry a mean to get control over the distance, or the time between two distant events, is provided by a laser signal exchanged between the bodies.
Accordingly, our basic measurement is just the time of flight. We thus consider a system of two particles exchanging a laser signal in the background of a plane gravitational wave propagating in the Minkowski spacetime, see Fig. \ref{fig:1}.
The wave is assumed to be moving with the speed ${\rm v}$ in $+z$ direction, so ${\rm h}_{\alpha\beta}[t,{\mathbf x}]:=h_{\alpha\beta}[u(t,{\mathbf x})]$, $u(t,{\mathbf x}):=t-\frac{{\mathbf \Omega}\cdot{\mathbf x}}{\rm v}$, with the unit vector ${\mathbf \Omega}=\left(0,0,1\right)$; we are going to consider all ten metric perturbations,
$h_{\alpha\beta}(u)$, 
separately.
We assume that observers, the emitter (E) and the detector (D), are moving freely in the underlying background; the equations of motion are derived from the Lagrange function
\begin{eqnarray}
 L & = & \frac12\left[ \left( -\dot{t}^2 + \dot{x}^2 + \dot{y}^2 + \dot{z}^2 + 
 {\rm h}_{\alpha\beta}\left(t, {\mathbf x}\right)\,\dot{x}^{\alpha}\,\dot{x}^{\beta} \right) \right].
\end{eqnarray}
In the above the over--dot `$\;\dot{\;}\;$` is used to denote derivative with respect to the proper time $\tau$ 
for trajectories of observers, $x_{E(D)}^{}(\tau)=[t_{E(D)}(\tau),{\mathbf x}_{E(D)}(\tau)]$,
and an affine parameter $\lambda$ for the light rays, $x_{ED}^{}(\lambda)=[t_{ED}(\lambda),{\mathbf x}_{ED}(\lambda)]$. 

In the remote past, before the arrival of the wave, motion of the observers can be characterized by their coordinate velocities ${\mathbf w}_{E}:=\frac{d{\mathbf x}_E}{dt}$ and ${\mathbf w}_{D}:=\frac{d{\mathbf x}_D}{dt}$. Thus the following form of the solution is considered ($x$ is a curve,
${\rm h}$ stands for a particular metric perturbation, '$\comp$' denotes composition of functions):
\begin{equation}
\label{eq:ansatz}
 \dot{t} = \gamma\left(1 + A^t\,{\rm h}\comp x^{}\right) \qquad
 \dot{\mathbf x} = \gamma\left({\mathbf w} + {\mathbf A}\, {\rm h}\comp x^{}\right), \qquad \gamma:=\frac{1}{\sqrt{1-{\mathbf w}^2}}, \qquad  {\mathbf w}^2=(w^x)^2+(w^y)^2+(w^z)^2<1
\end{equation}
with constants: ${\mathbf w}=\left(w^x,w^y,w^z\right)$, $A^t$ and ${\mathbf A}:=(A^x,A^y,A^z)$. 
Using the ansatz (\ref{eq:ansatz}) one can solve the equation of motion for the observers and express the constants $A^t$ and ${\mathbf A}$ in terms of the initial velocities ${\mathbf w}_{E}$ and ${\mathbf w}_{D}$; results are given in the Appendix \ref{sec:app}. For each perturbed metric, ${\rm g}_{\alpha\beta}=\eta_{\alpha\beta}+{\rm h}_{\alpha\beta}$, the trajectory $x^{}$ satisfies ${\rm g}_{\alpha\beta}\frac{d x^{\alpha}}{d\tau}\frac{d x^{\beta}}{d\tau}=-1 + {\rm O}(h^2)$. 

To the linear order in $h$ Eq. (\ref{eq:ansatz})  gives
\begin{equation}
\label{eq:eomED}
 \frac{d\tau}{dt} = \frac{1}{\gamma}\left(1 - A^t\,{\rm h}\comp x^{}\right),\qquad
 \frac{d{\mathbf x}}{dt} = {\mathbf w} + {\mathbf B}\, {\rm h}\comp x^{}, 
 \qquad {\mathbf B} = {\mathbf A} - A^t\,{\mathbf w}.
\end{equation}

Similarly, to find the light trajectory we use the ansatz:
\begin{equation*}
 \frac{dt}{d\lambda} = 1 + A_{L}^t\, {\rm h}\comp x^{}, \qquad \frac{d{\mathbf x}}{d\lambda} = {\mathbf n} + {\mathbf A}_{L}\,{\rm h}\comp x^{},
\end{equation*}
with constants: ${\mathbf n}:=(n^x,n^y,n^z)$, $n_x^2+n_y^2+n_z^2=1$, $A_{L}^t$ and ${\mathbf A}_{L}:=(A_{L}^x,A_{L}^y,A_{L}^z)$ (values of $A_{L}^t$ and ${\mathbf A}_{L}$ are given in the Appendix \ref{sec:app}).
In each case the solution satisfies ${\rm g}_{\alpha\beta} \frac{d x^{\alpha}}{d\lambda}\frac{d x^{\beta}}{d\lambda} = {\rm O}(h^2)$. Again, one obtains (to the linear order in $h$)
\begin{equation}
\label{eq:eomL}
 \frac{d\lambda}{dt} = 1 - A_{L}^t\,{\rm h}\comp x^{} \qquad 
 \frac{d {\mathbf x}}{dt} = {\mathbf n} + {\mathbf B}_{L}\,{\rm h}\comp x^{},\qquad {\mathbf B}_{L} = {\mathbf A}_{L} - A_{L}^t\,{\mathbf n}.
\end{equation}

\section{Time of flight}
\label{sec:dtau}

The time of flight is defined as the time difference between the emission of the laser signal at
${\mathbf x}_{\rm E}(t_E)$ and its reception at ${\mathbf x}_{\rm D}(t)$; it is given by
\begin{eqnarray}
\label{eq:tofd}
 \Delta\tau_{\rm ED}(t) &:= & \tau_{\rm D}(t) - \tau_{\rm E}(t_E) = \tau_{\rm D}(t) - \tau_{\rm E}(t-\delta t),
\end{eqnarray}
where $\tau_E$ and $\tau_D$ are the proper times of the emitter and the detector, $\delta t$ is the coordinate time interval between the two events. 

It is worth noting here a relation between the choice of initial conditions for the equations of motion (\ref{eq:eomED}), (\ref{eq:eomL}), and coordinate independence of the final result.
So far the coordinate system have been quite arbitrary, it was restricted only by the form of the metric perturbations, $h_{\alpha\beta}(u)$.
Thus as it stands, $\Delta\tau_{\rm ED}(t)$, may possibly define a coordinate--dependent quantity. This is so because any initial conditions set for the proper times of the observers in the form $\tau_{E}(t_0)=\tau_{E0}$, $\tau_{D}(t_0)=\tau_{D0}$ would tacitly assume synchronization of their clocks with respect to the coordinate time $t$. Consequently initial conditions would depend on the choice of coordinates. 
In general, one can choose a synchronization procedure at will (e.g. involving multiple exchange of the signals prior to the experiment) and try to incorporate it in a coordinate--independent way. Another possibility, adopted in the paper, is to single out a coordinate system in the past in accord to some chosen synchronization procedure and to treat it as a part of the underlying structure.
To this end we will assume that there are no perturbations in the past (say for $u\leq u_M$) and before that time the standard (Einstein) synchronization procedure was applied by an external observer, otherwise not participating in the experiment. Mathematically it can be achieved by selecting the Minkowski coordinates in the far past (say before $u_0$, $u_0<u_M$); moreover this choice of coordinates must be respected by all gauge transformations under consideration, i.e. condition of no "true`` nor ''coordinate`` plane waves in the remote past must be satisfied; see Sec. \ref{sec:gi}..

Initial conditions for the emitter, $x_E$, and detector, $x_D$, are therefore fixed at a time $t_0$: 
$$
\tau_E(t_0)=\tau_D(t_0)=\tau_0,
\qquad {\mathbf x}_E(t_0)={\mathbf x}_{E0},
\qquad {\mathbf x}_D(t_0)={\mathbf x}_{D0},
$$ 
and it is assumed that $u[t_0,{\mathbf x}_{D0}]<u_0<u_M$ and $u[t_0,{\mathbf x}_{E0}]<u_0<u_M$; this means that the proper times were set to $\tau_0$ at the coordinate time $t_0$ of the Minkowski coordinate system in the past, before the wave arrival, see Fig.\ref{fig:1}. 

Using Eq.(\ref{eq:eomED}) and (\ref{eq:tofd}) one gets for the time of flight (notation is explained in Appendix \ref{sec:app}):
\begin{eqnarray}
\label{eq:tau1}
 \Delta\tau_{\rm ED}(t) &= & (t-t_0)\left(\frac{1}{\gamma_D}-\frac{1}{\gamma_E}\right) - 
 \left[ \frac{A^t_D}{\gamma_D}\,\int\limits_{-\infty}^{t}\,{\rm h}[x_D(t')]\,dt' -
 \frac{A^t_E}{\gamma_E}\,\int\limits_{-\infty}^{t-\delta t}\,{\rm h}[x_E(t')]\,dt' \right]
 +\frac{\delta t}{\gamma_E}.
\end{eqnarray}
The time interval $\delta t$ can be determined from the equation of motion (\ref{eq:eomL}) for the light ray, $x_{ED}$, propagating from ${\mathbf x}_{\rm E}(t-\delta t)$ to ${\mathbf x}_{\rm D}(t)$:
\begin{equation}
 \label{eq:tau2}
 {\mathbf x}_{\rm E}(t-\delta t) + {\mathbf n}_{}\,\delta t + {\mathbf B}_{L}\,\int\limits_{0}^{\delta t}{\rm h}[x_{ED}(\lambda)]\,d\lambda = {\mathbf x}_D(t),
\end{equation}
where ${\mathbf n}_{}$ is an unit vector. It is understood that in the integrands of Eqs. (\ref{eq:tau1}), (\ref{eq:tau2}) the  zeroth order solutions (i.e. unperturbed trajectories)
\begin{equation}
\label{eq:up}
 x_{E}^{(0)}(t') = [t',\,{\mathbf x}_{E0} + {\mathbf w}_E(t'-t_0)], \qquad
 x_{D}^{(0)}(t') = [t',\,{\mathbf x}_{D0} + {\mathbf w}_D(t'-t_0)], \qquad
 x_{ED}^{(0)}(\lambda) = [t_E+\lambda,\,{\mathbf x}_{E0} + {\mathbf n}\,\lambda]
\end{equation}
have been taken. After splitting the time delay $\delta t=T+\delta T\;$ Eq.(\ref{eq:tau2}) can be solved iteratively with respect to small quantity $\delta T$. First iteration (for $h=0$) gives $T(t) = |{\mathbf x}_{D}^{(0)}(t) - {\mathbf x}_{E}^{(0)}(t-T(t))|$, which expresses the fact that $T(t)$ is the coordinate time interval between the emission and the detection for the unperturbed trajectories of the observers. From the second iteration one gets (to the linear order in $h$ and $\delta T$)
\begin{equation}
 \delta T = \frac{{\mathbf n}_0}{1-{\mathbf n}_0\cdot{\mathbf w}_E}\cdot
 \left( {\mathbf B}_D\,\int\limits_{-\infty}^{t}\,{\rm h}[x_{D}(t')]\,dt' - 
 {\mathbf B}_E\,\int\limits_{-\infty}^{t-T}\,{\rm h}[x_{E}(t')]\,dt' - 
 {\mathbf B}_{ED}\,\int\limits_{0}^{T}\,{\rm h}[x_{ED}(\lambda)]\,d \lambda\right),
\end{equation}
where ${\mathbf n}_0$ is the unit vector pointing from the detector to the emitter,
\begin{equation}
\label{eq:n0}
 {\mathbf n}_0(t):=\frac{{\mathbf x}_{D}^{(0)}(t) - {\mathbf x}_{E}^{(0)}[t-T(t)]}{T},
\end{equation}
${\mathbf B}_{E}$, ${\mathbf B}_{D}$ and ${\mathbf B}_{ED}$ are constants defined for the emitter, detector and light ray in the Appendix \ref{sec:app}.
This finally gives
\begin{eqnarray}
\label{eq:dtau}
 \Delta\tau_{\rm ED}(t) & = & 
 (t-t_0)\left(\frac{1}{\gamma_D}-\frac{1}{\gamma_E}\right) + \frac{T}{\gamma_E}
 - \left[ \frac{A^t_D}{\gamma_D}\,\int\limits_{-\infty}^{t}\,{\rm h}[x_{D}(t')]\,dt' -
 \frac{A^t_E}{\gamma_E}\,\int\limits_{-\infty}^{t-T}\,{\rm h}[x_{E}(t')]\,dt' \right] 
 \nonumber\\
 && + \frac{{\mathbf n}_0}{\gamma_E(1-{\mathbf n}_0\cdot{\mathbf w}_E)}\cdot
 \left( {\mathbf B}_D\,\int\limits_{-\infty}^{t}\,{\rm h}[x_{D}(t')]\,dt' - 
 {\mathbf B}_E\,\int\limits_{-\infty}^{t-T}\,{\rm h}[x_{D}(t')]\,dt' - 
 {\mathbf B}_{ED}\,\int\limits_{0}^{T}\,{\rm h}[x_{ED}(\lambda)]\,d \lambda\right).
\end{eqnarray}
In the Eq.(\ref{eq:dtau}) all integrals are taken along the unperturbed trajectories $(\ref{eq:up})$; in addition for the light path, $x^{(0)}_{ED}$, one can assume $t_E=t-T$ and ${\mathbf n}={\mathbf n}_0$. Substitution of the constants $A^t$, ${\mathbf A}$, ${\mathbf B}$, given in the Appendix \ref{sec:app} into the Eq.(\ref{eq:dtau}) completes the derivation of the time of flight for each particular mode. For a linear combination of modes the response consists of the perturbation--free part of (\ref{eq:dtau}) and the superposition of the corresponding $h$--dependent terms. 

Thus far we have considered a wave moving along the $z$ direction. In consequence 
the final formula, Eq.(\ref{eq:dtau}), is valid in
an orthonormal basis $\{{\mathbf e}_{sx}\sim\partial_x, {\mathbf e}_{sy}\sim\partial_y,{\mathbf e}_{sz}\equiv{\mathbf \Omega}\}$ (we call it the source frame). However one is interested in a more general situation when a wave is moving in an arbitrary direction ${\mathbf \Omega}$. To obtain the result valid in an arbitrary orthonormal frame, $\{{\mathbf e}_x,{\mathbf e}_y,{\mathbf e}_z\}$, related to the source frame by an orthogonal matrix ${\mathbf R}$, $R_{ij}={\mathbf e}_{i}\cdot{\mathbf e}_{sj}$ one should make the following substitutions in the original formula (\ref{eq:dtau}):
\begin{eqnarray}
 A^t_E({\mathbf w}^{s}_E) & \rightarrow & A^t_E({\mathbf R}^T\cdot{\mathbf w}_E)\nonumber\\
 A^t_D({\mathbf w}^{s}_D) & \rightarrow & A^t_D({\mathbf R}^T\cdot{\mathbf w}_D)\nonumber\\
 {\mathbf B}_E({\mathbf w}^{s}_E) & \rightarrow & {\mathbf R}\cdot{\mathbf B}_E({\mathbf R}^T\cdot{\mathbf w}_E) \nonumber\\
 {\mathbf B}_D({\mathbf w}^{s}_D) & \rightarrow & {\mathbf R}\cdot{\mathbf B}_D({\mathbf R}^T\cdot{\mathbf w}_D) \nonumber\\
 {\mathbf B}_{ED}({\mathbf n}^{s}_0) & \rightarrow & {\mathbf R}\cdot{\mathbf B}_{ED}({\mathbf R}^T\cdot{\mathbf n}_0). \nonumber
\end{eqnarray}
In the above ${\mathbf w}^{s}_E$, ${\mathbf w}^{s}_D$, ${\mathbf n}^{s}_0$ refer to components in the source frame; ${\mathbf w}_E$, ${\mathbf w}_D$, ${\mathbf n}_0$ are components in the new basis; dot denotes matrix multiplication of a column vector.
%

\section{Numerical simulations}
\label{sec:num}

As an example we consider a scalar monochromatic gravitational wave 
$h(u) = H\,\Theta(u-u_M)\,\cos{[2\pi f(u-u_M)]}$ with the phase $u=t-\frac{z}{\rm v}$, ${\rm v}=0.6$, the amplitude $H=0.09$ and the frequency $f=0.1$ Hz originating at $u_M=0$; $\Theta(x)$ is the Heaviside step function. Scalar wave is defined here by the the requirement that in a preferred (conformal) coordinates it has the form
$$
{\rm g}_{\mu\nu}=\left(1+h\right)\,\eta_{\mu\nu}.
$$
The approach presented in the paper allows for the use a restricted class of almost Minkowskian coordinates (that is the class of coordinates in which the metric perturbations have the form $h_{\alpha\beta}[u(t,{\mathbf x})]$) to address the issue (for the full account of the gauge invariance we refer the reader to sec. \ref{sec:gi}). The advantage of the conformal coordinate system however relies on the simplicity of the null trajectories which remain undisturbed in this gauge, so the numerical studies of the response can easily be carried out. To obtain the time of flight in the conformal coordinates one has to consider the superposition of four responses (their $h$--dependent parts) given by Eq.(\ref{eq:dtau}). One can check that, indeed, in this case the coefficients ${\mathbf B}_{ED}$ at the last term add up to zero, and the net effect arises solely due to the disturbed motion and modified proper times of the observers. 

\begin{figure}[htp]
\begin{center}
\includegraphics[width=22pc]{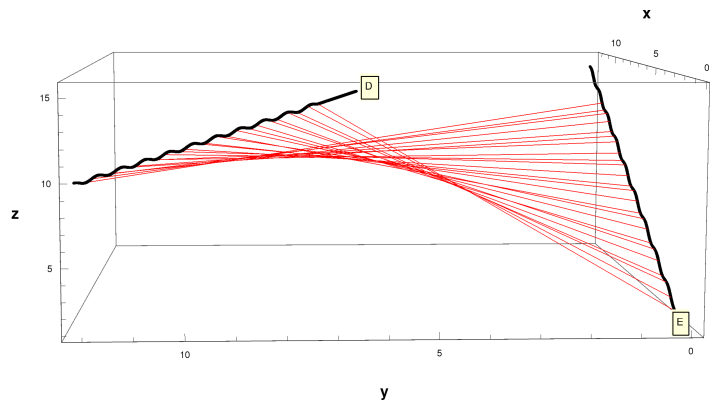}
\end{center}
\caption{Trajectories of the emitter (E), detector (D) and the exchanged light signals (red)
for the monochromatic scalar wave; the assumed time interval: $0<t<100\,$sec. Framed labels ''E" and ``D" mark the positions at $t=0$; other details are given in the text.}
\label{f:2}
\end{figure}

\begin{figure}[htp]
\begin{center}\includegraphics[width=26pc]{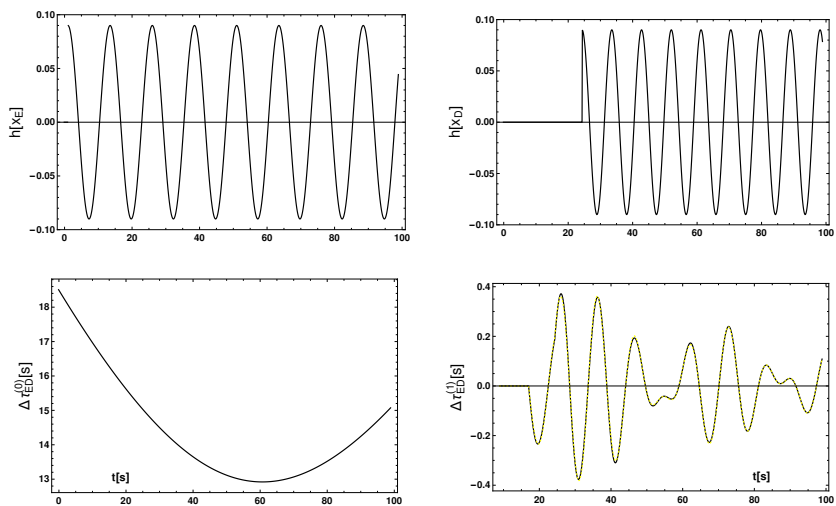}
\end{center}
\caption{Upper plots: gravitational wave amplitude, $h[u(t,{\mathbf x}(t))]$, at the emitter (left) and at the detector (right) as functions of the coordinate time. One can observe the time delay of the wave arrival to the detector and the gravitational Doppler effect (change of the frequency). Lower plots: emitter--detector time delay for the unperturbed trajectories,
$\Delta\tau_{ED}^{(0)}$ (left), and contribution due to the perturbations, $\Delta\tau_{ED}^{(1)}:=\Delta\tau_{ED}-\Delta\tau_{ED}^{(0)}$ (right). On the right diagram one can observe that the numerical (black, solid line) and the analytical (yellow, dotted line) responses are in good agreement.}
\label{f:3}
\end{figure}

The space diagram of the emitter, detector and light trajectories is displayed in Fig. \ref{f:2}. The clocks are synchronized to $\tau_{E0}=\tau_{D0}=t_0=-4\,$sec.; the assumed initial positions (in sec.) and the velocities read ${\mathbf x}_{E0} = (0,0,0)$,  ${\mathbf x}_{D0} = (0,6,16)$, $\;{\mathbf w}_{E} = (0.1,\,0,\,0.12)$, ${\mathbf w}_{D} = (0,\,0.05,\,-0.05)$.
Lower diagrams in Fig. \ref{f:3} show the time of flight for the unperturbed Minkowski spacetime, 
$\Delta\tau_{ED}^{(0)}$, and the contribution due to the perturbations, i.e. $\Delta\tau_{ED}^{(1)}:=\Delta\tau_{ED}-\Delta\tau_{ED}^{(0)}$, as derived from Eq. (\ref{eq:dtau}). The results of the semi-numerical simulations are also presented. To carry them out one first solves the equations of motion (\ref{eq:eomED}) and obtains the perturbed trajectories ${\mathbf x}_{E}(t)$ and ${\mathbf x}_{D}(t)$; the coordinate time delays $\delta t(t)$ (needed to compute $\Delta\tau_{ED}$) are then found iteratively (to an assumed precision) from the equation $\delta t(t)=|{\mathbf x}_{D}(t)-{\mathbf x}_{E}[t-\delta t(t)]|$.

One can also compare the results obtained in different coordinates. For example, the gauge transformations generated by ${\boldsymbol \zeta}=\left( \zeta,0,0,-\frac{1}{\rm v}\zeta \right)$, with $\zeta(u)=-\frac12\,\int\limits_{-\infty}^{u}h(u')\,du'$ lead to the metric (\cite{AB2015})
\begin{equation}
\label{eq:mTT}
{\rm g}'_{\mu\nu}=\eta_{\mu\nu} + h\,\left(dx^2+dy^2\right)+\left(1-\frac{1}{{\rm v}^2}\right)h\,dz^2.
\end{equation}
The time of flight for perturbations (\ref{eq:mTT}) receives contributions from all terms entering the formula (\ref{eq:dtau}), including the one along the light trajectory; it perfectly agrees with numerical simulations and the response obtained in the conformal gauge.

\section{Special cases}

\subsection{$[h]=0$}

When there is no wave (i.e. for the flat spacetime) the response (\ref{eq:dtau}) reduces to the standard special--relativistic Doppler effect expressed in arbitrary nearly--flat coordinates. In this case the $h$--dependent part of (\ref{eq:dtau}) (interpreted as arising from ''coordinate waves'') is zero. To check this one can make use of gauge freedom and choose a specific coordinate system (gauge--invariance is addressed in sec. \ref{sec:gi}). In particular, when the Minkowski coordinates are selected one obtains
\begin{eqnarray}
 \Delta\tau_{ED} & = & (t-t_0)\left(\frac{1}{\gamma_D}-\frac{1}{\gamma_E}\right)+\frac{T(t)}{\gamma_E} \\
 \frac{d\tau_E}{d\tau_D} & = & 1-\gamma_D\,\frac{d}{dt}\Delta\tau_{ED} = 
 1 - \gamma_D\,\left( \frac{1}{\gamma_D}-\frac{1}{\gamma_E} + \frac{1}{\gamma_E}\frac{d\,T}{d\,t} \right) \nonumber \\
 \label{eq:dtdt}
 & = & \frac{\gamma_D}{\gamma_E}
 \frac{1-{\mathbf n}_0\cdot{\mathbf w}_D}{1-{\mathbf n}_0\cdot{\mathbf w}_E} = 
 \frac{{\mathbf U}_D\cdot{\mathbf \Lambda}}{{\mathbf U}_E\cdot{\mathbf \Lambda}},
\end{eqnarray}
where ${\mathbf U}_{E(D)}=\frac{\partial}{\partial\tau}_{\Bigr|_{E(D)}}$ and ${\mathbf \Lambda}=\frac{\partial}{\partial\lambda}$ are the four-velocity vectors of the observers and the light ray, respectively. In the second line of Eq.(\ref{eq:dtdt}) we used the fact that the time derivative of the unperturbed time delay can be expressed as $\frac{dT}{dt}=\frac{{\mathbf n}_0\cdot\left({\mathbf w}_D-{\mathbf w}_E\right)}{1-{\mathbf n}_0\cdot{\mathbf w}_E}$; this relation follows from Eq.(\ref{eq:n0}).

\subsection{${\mathbf w}_{E}={\mathbf w}_{D}$}

\label{sec:w}

When the observers are keeping a constant distance, i.e. when ${\mathbf w}_{E}={\mathbf w}_{D}=:{\mathbf w}$ the response can be rewritten in the following compact form
\begin{eqnarray}
\label{eq:equi}
 \Delta\tau_{\rm ED} & = & 
 \frac{T}{\gamma} + \frac{1}{\gamma(1-{\mathbf w}\cdot{\mathbf n}_0)}\,\left[ \frac{1-\frac{{\mathbf \Omega}\cdot{\mathbf n}_0}{\rm v}}{1-\frac{{\mathbf \Omega}\cdot{\mathbf w}}{\rm v}} 
 \left( {\mathbf A}\cdot{\mathbf n}_0 -\tilde{A}^t \right)-{\mathbf B}_{ED}\cdot{\mathbf n}_0\right]\,\int\limits_{0}^{T}{\rm h}[x_{ED}(\lambda)]\,d\lambda\nonumber\\
 &=&
 \frac{T}{\gamma} - \frac{{\mathbf n}_{\mathbf w}\otimes{\mathbf n}_{\mathbf w}}{2\,{\mathbf U}\cdot{\mathbf \Lambda}}:{\mathbf s}\,\int\limits_{0}^{T}{\rm h}[x_{ED}(\lambda)]\,d\lambda,
\end{eqnarray}
where ${\mathbf n}_{\mathbf w}:={\mathbf n}_0-\gamma\,\frac{{\mathbf \Lambda}\cdot{\mathbf K}}{{\mathbf U}\cdot{\mathbf K}}\,{\mathbf w}$, $K^{\alpha}\equiv-\nabla^{\alpha}\,u=\left(1,\frac{\mathbf \Omega}{\rm v}\right)$. 
In the above equation and in what follows the colon denotes tensor contraction, e.g. $A\otimes B:C=A^i\,B^j\,C_{ij}$, etc.
In the formula (\ref{eq:equi}) the tensors ${\mathbf s}$ are playing the role of the (modified) wave polarization tensors. In the 
source frame they are given by
\begin{equation}
\begin{array}{rl}
 h_{tt} &: \qquad {\mathbf s} = \frac{{\boldsymbol \Omega}\otimes{\boldsymbol \Omega}}{{\rm v}^2} \\ &\\
 h_{ti}+h_{it} &: \qquad 
 {\mathbf s} = \frac{{\boldsymbol \Omega}\otimes{\mathbf e}_{si} + {\mathbf e}_{si}\otimes{\boldsymbol \Omega}}{{\rm v}} \\ &\\
 h_{ij}+h_{ji} &: \qquad
 {\mathbf s} = {\boldsymbol e}_{si}\otimes{\boldsymbol e}_{sj} + {\boldsymbol e}_{sj}\otimes{\boldsymbol e}_{si} \\ &\\
\end{array}
\end{equation}

Instead of referring to the time of flight for a particular signal one can also consider observations of the time rates of the incoming signals (the Doppler tracking experiment). In this case the first and second $\tau_D$-derivatives of $\tau_{ED}$ can serve as observables. For equidistant observers the unit vector ${\mathbf n}_0$ and the unperturbed time delay $T$ are constant therefore the time derivatives of $\Delta\tau_{ED}$ (i.e. Doppler observables) can be easily obtained; they read:
\begin{eqnarray}
\label{eq:dtaup}
 \frac{d}{d\tau}\,\Delta\tau_{\rm ED} & = & -\frac12\,\frac{\left({\mathbf n}_{\mathbf w}\otimes{\mathbf n}_{\mathbf w}\right):{\mathbf s}}{{\mathbf U}\cdot{\mathbf \Lambda}}\,\gamma\left(1-\frac{{\mathbf \Omega}\cdot{\mathbf w}}{\rm v}\right)\,\int\limits_{0}^{T}h'[u(\lambda)]\,d\lambda \\
 \label{eq:dtaupp}
 \frac{d^2}{d\tau^2}\,\Delta\tau_{\rm ED} & = & -\frac12\,\frac{\left({\mathbf n}_{\mathbf w}\otimes{\mathbf n}_{\mathbf w}\right):{\mathbf s}}{{\mathbf U}\cdot{\mathbf \Lambda}}\,\gamma^2\left(1-\frac{{\mathbf \Omega}\cdot{\mathbf w}}{\rm v}\right)^2\,\int\limits_{0}^{T}h''[u(\lambda)]\,d\lambda
\end{eqnarray}
where we have shortened the notation in the integrals: $u(\lambda)\equiv u[x_{ED}(\lambda)]$.

Few comments are in order.
 In obtaining the time of flight in the form given in (\ref{eq:equi}), the integrals along the world lines of the observers that determined their proper times were transformed to the integral along the null trajectory of the laser light. 
 This simplifies the result and imposes an interpretation that the time of flight depends only on the perturbations along the path of the light between the emitter and detector. The independence on the past history (i.e. the state the wave prior to $u_E$) for the equidistant observers can arise because their clocks and world lines are equally affected by the wave prior to $u_E$.
 For observers moving with different velocities the response $\Delta\tau_{ED}$ can be sensitive to the past history of the emitter and the detector. For a general gravitational field (as opposed to a plane gravitational wave) the time of flight can be sensitive to the past history
 even when ${\mathbf w}_E={\mathbf w}_D$; this is because in this case gravitational field can affect world lines of the emitter, detector and light \emph{independently}
 
 It can be checked by lengthy, but direct calculations that the detector response (\ref{eq:equi}) can be written in an explicitly gauge--invariant form as
\begin{eqnarray}
\label{eq:ginv}
 \Delta\tau_{\rm ED} & = & \frac{T}{\gamma} + \frac{1}{{\mathbf U}\cdot{\mathbf \Lambda}}\,{\mathbf U}\otimes{\mathbf \Lambda}\otimes{\mathbf U}\otimes{\mathbf \Lambda}\,:\,
 \int\limits_{0}^{T}\, {\mathbf R}^{(1)}\left[k(u(\lambda))\right]\,d\lambda ,
\end{eqnarray}
where, as before, ${\mathbf U}$ is the four--velocity of the emitter or detector, and 
${\mathbf \Lambda}=\frac{\partial}{\partial {\mathbf \lambda}}$ is the four--velocity of the light ray w.r.t. the affine parameter $\lambda$ normalized as in Eq.(\ref{eq:eomED}).
In Eq. (\ref{eq:ginv}) ${\mathbf R}^{(1)}$ is the linear part of the Riemann tensor (with components $2R^{(1)}_{\alpha\mu\beta\nu}=
\partial_{\mu\beta}h_{\alpha\nu}+\partial_{\nu\alpha}h_{\beta\mu}-\partial_{\alpha\beta}h_{\mu\nu}-\partial_{\mu\nu}h_{\alpha\beta}$)
and $k$ is related to $h$ by the condition
\begin{eqnarray}
\frac{d^2}{d\,u^2}\,k(u) = h(u).
\end{eqnarray}
It follows then that the second time derivative of the time of flight is given by (\cite{KF14})
\begin{eqnarray}
\label{eq:d2Dtau}
 \frac{d^2}{d\,\tau^2}\,\Delta\tau_{\rm ED} & = & 
 \frac{1}{{\mathbf U}\cdot{\mathbf \Lambda}}\,{\mathbf U}\otimes{\mathbf \Lambda}\otimes{\mathbf U}\otimes{\mathbf \Lambda}\,:\,
 \int\limits_{0}^{T}\, {\mathbf R}^{(1)}\left[h(u(\lambda))\right]\,d\lambda.
\end{eqnarray}


\subsection{${\mathbf w}_{E}={\mathbf w}_{D}=0$, ${\rm v}=1$}

When the emitter and detector are at rest and the wave moves with speed of light the responses reduce to
\begin{eqnarray}
\label{eq:w0}
 \Delta\tau_{\rm ED} & = & T + \frac{{\mathbf n}_0\otimes{\mathbf n}_0:{\mathbf s}}{2}\,\int\limits_{0}^{T}h[u(\lambda)]\,d\lambda
\end{eqnarray}
with the polarization tensors
\begin{equation}
\begin{array}{rl}
 h_{tt} &: \qquad {\mathbf s} = {\boldsymbol \Omega}\otimes{\boldsymbol \Omega} \\ &\\
 h_{ti}+h_{it} &: \qquad 
 {\mathbf s} = {\boldsymbol \Omega}\otimes{\mathbf e}_{si} + {\mathbf e}_{si}\otimes{\boldsymbol \Omega} \\ &\\
 h_{ij}+h_{ji} &: \qquad
 {\mathbf s} = {\boldsymbol e}_{si}\otimes{\boldsymbol e}_{sj} + {\boldsymbol e}_{sj}\otimes{\boldsymbol e}_{si} \\ &\\
\end{array}
\end{equation}

\section{Gauge invariance}

\label{sec:gi}
In deriving the response a particular coordinate system have been used, so the final formula for the time of flight represents seemingly coordinate dependent quantity. Thus, ultimately, it should be checked whether the Eq.(\ref{eq:dtau}) actually defines an observable. We explicitly show that indeed this is the case, and the time of flight $\Delta\tau_{ED}$ is gauge invariant in the following (usual) sense. Each response is valid not only for a particular wave ${\rm h}_{\alpha\beta}$ and trajectories $x_E$, $x_D$ and $x_{ED}$, but is unambiguously defined for the equivalence class of this system defined as
\begin{eqnarray}
\label{eq:gi}
\left[ {\rm h}_{\alpha\beta},\,x_E,\, x_D,\,x_{ED} \right] & := & 
\left\{ {\rm h}_{\alpha\beta} + \mathsterling_{\xi}{\rm h}_{\alpha\beta},\,
x_{E} + \mathsterling_{\xi}x_E,\,
x_{D} + \mathsterling_{\xi}x_D,\,
x_{ED} + \mathsterling_{\xi}x_{ED};\right.\quad \nonumber \\
&&
\qquad\qquad\;
\left.\xi^{\alpha}(t,{\mathbf x})=\zeta^{\alpha}\left(t-\frac{{\mathbf \Omega}\cdot{\mathbf x}}{\rm v}\right),\;
{\mathbf \Omega}=(0,0,1),\;
\;\zeta^{\alpha}\sim{\rm O}(h)\right\}
\end{eqnarray}
where the Lie derivatives with respect to a vector field $\xi^{\alpha}$ are given by
$$
\mathsterling_{\xi}{\rm h}_{\alpha\beta}=\xi_{\alpha,\beta}+\xi_{\beta,\alpha},\quad
\mathsterling_{\xi}x^{\alpha}_E = \xi^{\alpha}\comp x_E,\quad
\mathsterling_{\xi}x_D = \xi^{\alpha}\comp x_D,\quad
\mathsterling_{\xi}x_{ED} = \xi^{\alpha}\comp x_{ED}.
$$
It is further assumed that the generators $\zeta^{\alpha}$ in (\ref{eq:gi}) vanish in a far past; this amounts to saying that the standard synchronization procedure of the clocks was adopted once and for all before the wave arrival.

First we consider the gauge independence for $\zeta^{\alpha}=\left(\eta,0,0,0)\right)$ generator.
Transformations of the metric 
$$
{\rm h}_{tt} \xrightarrow{} {\rm h}_{tt} - 2\,\xi_{,t},\quad {\rm h}_{tz} \xrightarrow{} {\rm h}_{tz} - \xi_{,z},
$$
change the integrals of Eq.(\ref{eq:dtau}) in the following way:
\begin{eqnarray*}
 \int\limits_{-\infty}^{t-T}\,\mathsterling_{\xi}h[u_{E}(t')]\,dt' & = & \left(-2 + \frac{1}{\rm v}\right)\,
 \frac{1}{1-\frac{{\mathbf \Omega}\cdot{\mathbf w}_{E}}{\rm v}}\,\eta\left(u_E\right)\\
 \int\limits_{-\infty}^{t}\, \mathsterling_{\xi}h[u_{D}(t')]\,dt' & = & \left(-2 + \frac{1}{\rm v}\right)\,
 \frac{1}{1-\frac{{\mathbf \Omega}\cdot{\mathbf w}_{D}}{\rm v}}\,\eta\left(u_D\right)\\
 \int\limits_{0}^{T}\, \mathsterling_{\xi}h[u_{ED}(t')]\,dt' & = & \left(-2 + \frac{1}{\rm v}\right)\,
 \frac{1}{1-\frac{{\mathbf \Omega}\cdot{\mathbf n}_{0}}{\rm v}}\,\left[\eta\left(u_D\right)-\eta\left(u_E\right)\right],
 \end{eqnarray*}
whereas the transformations of the trajectories
$$
x^{\alpha}_{E} \rightarrow x^{\alpha}_{E} + \delta^{\alpha}_{t}\,\eta[u(x_E)],\quad
x^{\alpha}_{D} \rightarrow x^{\alpha}_{D} + \delta^{\alpha}_{t}\,\eta[u(x_D)],\quad
x^{\alpha}_{ED} \rightarrow x^{\alpha}_{ED} + \delta^{\alpha}_{t}\,\eta[u(x_{ED})],
$$
lead to
\begin{eqnarray}
&&
\tau_E(t)\rightarrow\tau_E(t)-\frac{1}{\gamma_E}\,\eta[u(x_E(t))],\quad
\tau_D(t)\rightarrow\tau_D(t)-\frac{1}{\gamma_D}\,\eta[u(x_D(t))],
\label{eq:tr}\\
&&
{\mathbf x}_{E}(t)\rightarrow {\mathbf x}_{E}(t) - {\mathbf w}_{E}\,\eta[u(x_{E}(t))],\quad
{\mathbf x}_{D}(t)\rightarrow {\mathbf x}_{D}(t) - {\mathbf w}_{D}\,\eta[u(x_{D}(t))],\quad
{\mathbf x}_{ED}(t)\rightarrow {\mathbf x}_{ED}(t) - {\mathbf n}_0\,\eta[u(x_{ED}(t))]\nonumber.
\end{eqnarray}
Transformations (\ref{eq:tr}) modify the $h$--independent term of $\Delta\tau_{ED}$; they amount to adding
\begin{eqnarray*}
 \frac{1}{\gamma_D}\eta(u_D)-\frac{1}{\gamma_E}\eta(u_E)+\frac{1}{\gamma_{E}
 \left(1-{\mathbf n}_0\cdot{\mathbf w}_E \right)}
 \left\{ {\mathbf n}_0\cdot{\mathbf w}_D\,\eta(u_D)-{\mathbf n}_0\cdot{\mathbf w}_E\,\eta(u_E)
 -\left[ \eta(u_D) - \eta(u_E)\right] \right\}
\end{eqnarray*}
to the response. Lengthy, but direct calculations that make use of the results given in the Appendix \ref{sec:app} now show that the combined transformations for the metric and the trajectories leave the response (\ref{eq:dtau}) unchanged. Following along the above example one can demonstrate the gauge independence for other generators: $\left(0,\eta,0,0)\right)$, $\left(0,0,\eta,0)\right)$ and $\left(0,0,0,\eta\right)$; one can note that in these three cases $\Delta\tau_{ED}$ is invariant separately with respect to the transformations of the metric and the transformations of the paths.

\section{Conclusions}

In the paper we have considered the time of flight of the light signal exchanged between two freely, but independently moving observers as the basic observable of the interferometric detectors for a plane gravitational wave moving with an arbitrary velocity. The coordinate dependent approach was adopted (as opposed to a geometric one) but no particular gauge was assumed. Consequently, the final result to represent a truly measurable quantity must have been proved to be coordinate independent (gauge invariant). It is clear from the experimental context that it must be the case, but nevertheless the invariance (on the perturbative level) was explicitly demonstrated. Although this is not a standard approach in the field of gravitational wave detection, it is conceptually strongly embedded in all geometric theories of gravity (GR included). 

The result is applicable not only in the context of laser interferometry but, in a natural way, to the pulsar timing method of detection as well. 
It may be used in gravitational wave data analysis in cases where the standard gauges (Lorentz, TT, etc.)
are not achievable or for more complex signals comprising polarization modes moving with different velocities as is predicted by some alternative theories of gravity. It also enables extensions: it may be applied to a general (weak) gravitational field in the flat spacetime, or in the case of gravitational waves in the cosmological background \cite{AB18}.

\section{Acknowledgments}

The work was supported in part by the National Science Centre grant UMO-2017/26/M/ST9/00978.

\appendix

\section{Solutions of the equations of motion}

\label{sec:app}

Values of $A^t({\mathbf w})$ and ${\mathbf A}({\mathbf w})$ determined from the equation of motion (\ref{eq:eomED}), (\ref{eq:eomL}) for the observers and the light ray:
$$
\begin{array}{c||llllll}
 h_{tt}(u)\,\dot{t}^2 & \quad A^t = 1+\frac12\frac{\rm v}{w^z-{\rm v}}, & {\mathbf A} = \left(0,0,\frac12\frac{1}{w^z-{\rm v}}\right) \\
 2\,h_{tx}(u)\,\dot{t}\,\dot{x} & \quad A^t = \frac{w^x\,w^z}{w^z-{\rm v}}, & {\mathbf A} = \left(-1,0,\frac{w^x}{w^z-{\rm v}}\right) &  \\
 2\,h_{ty}(u)\,\dot{t}\,\dot{y} & \quad A^t = \frac{w^y\,w^z}{w^z-{\rm v}}, & {\mathbf A} = \left(0,-1,\frac{w^y}{w^z-{\rm v}}\right) & \\ 
 2\,h_{tz}(u)\,\dot{t}\,\dot{z} & \quad A^t = \frac{(w^z)^2}{w^z-{\rm v}}, & {\mathbf A} = \left(0,0,\frac{\rm v}{w^z-{\rm v}}\right) & \\
 h_{xx}(u)\,\dot{x}^2 & \quad A^t = \frac12\frac{(w^x)^2}{w^z-{\rm v}}{\rm v}, & {\mathbf A} = \left(-w^x,0,\frac12\frac{(w^x)^2}{w^z-{\rm v}}\right) & \\
 h_{yy}(u)\,\dot{y}^2 & \quad A^t = \frac12\frac{(w^y)^2}{w^z-{\rm v}}{\rm v}, & {\mathbf A} = \left(-0,w^y,\frac12\frac{(w^y)^2}{w^z-{\rm v}}\right) & \\
 h_{zz}(u)\,\dot{z}^2 & \quad
 A^t = \frac12\frac{(w^z)^2}{w^z-{\rm v}}{\rm v}, & {\mathbf A} = \left(0,0,-w^z+\frac12\frac{(w^z)^2}{w^z-{\rm v}}\right) & \\
 2\,h_{xy}(u)\,\dot{x}\,\dot{y} & \quad A^t = \frac{w^x\,w^y}{w^z-{\rm v}}{\rm v}, & {\mathbf A} = \left(-w^y,-w^x,\frac{w^x\,w^y}{w^z-{\rm v}}\right) & \\
 2\,h_{xz}(u)\,\dot{x}\,\dot{z} & \quad A^t = \frac{w^x\,w^z}{w^z-{\rm v}}{\rm v}, & {\mathbf A} = \left(-w^z,0,\frac{w^x}{w^z-{\rm v}}{\rm v}\right) & \\
 2\,h_{yz}(u)\,\dot{y}\,\dot{z} & \quad A^t = \frac{w^y\,w^z}{w^z-{\rm v}}{\rm v}, & {\mathbf A} = \left(0,-w^z,\frac{w^y}{w^z-{\rm v}}{\rm v}\right) & \\
\end{array}
$$
For the light ray one has: $A^t_{L}=A^t({\mathbf n})$, 
${\mathbf A}_{L}={\mathbf A}({\mathbf n})$. 

The following notation is used in the paper: for the emitter, 
$A^t_{E}:=A^t({\mathbf w}_{E})$, 
${\mathbf A}_{E}:={\mathbf A}({\mathbf w}_{E})$, ${\mathbf B}_{E}:={\mathbf A}_{E}-A^t_{E}\,{\mathbf w}_{E}$;
for the detector,
$A^t_{D}:=A^t({\mathbf w}_{D})$, 
${\mathbf A}_{D}:={\mathbf A}({\mathbf w}_{D})$, ${\mathbf B}_{D}:={\mathbf A}_{D}-A^t_{D}\,{\mathbf w}_{D}$;
for the light trajectory, ${\mathbf B}_{ED}:={\mathbf A}({\mathbf n}_0)-A^t({\mathbf n}_0)\,{\mathbf n}_0$ 



%
%

\begin{thebibliography}{99}
%
\bibitem{MTW} Ch. W. Misner, K. S. Thorne, J. A. Wheeler, {\it Gravitation},
W. H. Freeman and Company, San Francisco (1973).
%
\bibitem{AdLIGO} www.ligo.org.
%
\bibitem{LG77} L. Grishchuck, Sov. Phys. --Usp {\bf 20}, (1977).
%
\bibitem{GP80} L. Grishchuck, A. G. Polnarev, {\it General Relativity and gravitation: One hundred years after the birth of Albert Einstein}, v{\bf 2}, A. Held, New York: Plenum, (1980).
%
\bibitem{BG04} D. Baskaran, L. Grishchuck, Classical and Quantum Gravity {\bf 21}, 4041 (2004).
%
\bibitem{MR14} M. Rakhmanov, Classical and Quantum Gravity {\bf 31}, 085006 (2014).
%
\bibitem{ET} \url{https://www.et-gw.eu}.
%
\bibitem{LISA} \url{https://www.lisamission.org}.
%
%
\bibitem{WillBook} C. M. Will, {\it Theory and Experiment in Gravitational Physics},
Cambrige University Press, Cambrige, England (1993).
%
\bibitem{CFPS2012} T. Clifton, P. G. Ferreira, A. Padilla, C. Skordis, Phys. Rep. {\bf 513}, 1 (2012).
%
\bibitem{FM2003} Y. Fujii, K. Maeda, {\it The Scalar--Tensor Theory of Gravitation},
Cambrige University Press, Cambrige, England (2003).
%
S. Capozziello, M. De Laurentis , Phys. Rep. {\bf 509}, 167 (2011);
%
\bibitem{SF2010} T. P. Sotiriou, V. Faraoni, Rev. Mod. Phys. {\bf 82}, 451 (2010).
%
\bibitem{YS13} N. Yunes, X. Siemens, Living Rev. Relativ. 16: 9 (2013).
%
\bibitem{Eardley73} D. M. Eardley, D. L. Lee, A. P. Lightmann, Phys. Rev. D {\bf 8}, 3308 (1973).
%
\bibitem{PWBook}  E. Poisson, C. M. Will {\it Gravity. Newtonian, Post-Newtonian, Relativistic},
Cambrige University Press (2014).
%
\bibitem{CYC12} K. Chatziioannou, N. Yunes, N. Cornish, Phys. Rev. D {\bf 86}, 022004 (2012).
%
\bibitem{HN13} K. Hayama, A. Nishizawa, Phys. Rev. D {\bf 87}, 062003 (2013).
%
\bibitem{TA10} M. Tinto, M. E. D. S. Alves, Phys.Rev. {\bf D 82}, 122003 (2010).
%
\bibitem{AB2012} A.B{\l}aut, Phys. Rev. D {\bf 85}, 043005 (2012).
%
\bibitem{IWMP15} M. Isi, A. J. Weinstein, C. Mead, M. Pitkin,  arXiv:1502.00333.
%
\bibitem{Jac07} T. Jacobson, Phys. Rev. Lett. {\bf 83}, 2699 (1999); 
T. Jacobson, D. Mattingly, Phys.Rev. D {\bf 70}, 024003 (2004); 
T. Jacobson, PoS QG-PH (2007), 020,
From Quantum To Emergent Gravity: Theory and Phenomenology,
June 11-15 2007, SISSA; Trieste Italy; arxiv:0801.1547v2; 
%
\bibitem{YBBY17} K. Yagi, D. Blas, E. Barausse, N. Yunes, Phys. Rev. D {\bf 89}, 084067 (2014).
%
\bibitem{BCL10} C. Bogdanos, S. Capozziello, M. De Laurentis, Astropart. Phys. {\bf 34}, 236-244 (2010). 
%
\bibitem{CST10} S..  Capoziello, A. Stabile, A. Troisi, Int. J. Theor. Phys. {\bf 49}, 1251-1261 (2010).
%
\bibitem{MagNic00} M. Maggiore, A. Nicolis, Phys. Rev. D {\bf 62}, 024004 (2000).
%
\bibitem{Nakao01} K. I. Nakao, T. Harada, M. Shibata, S. Kawamura, T. Nakamura,
Phys. Rev. D, {\bf 63}, 082001 (2001).
%
\bibitem{AB2015} A. B{\l}aut, BPhys. Rev. D 92, 063013 (2015).
%
\bibitem{KF14} M. J. Koop, L. S. Finn, Phys. Rev. D {\bf 90}, 062002 (2014).
%
\bibitem{AB18} A. B{\l}aut, in preparation.
\end{thebibliography}
\end{document}